\documentclass[prl, aps,  twocolumn,reprint,
showpacs]{revtex4-1}

\usepackage{graphicx}
\usepackage{dcolumn}
\usepackage{bm}

\usepackage[colorlinks,urlcolor=blue,citecolor=blue,linkcolor=blue]{hyperref}

\usepackage{comment}
\usepackage{color}
\usepackage{physics}
\usepackage{cancel}

\newcommand{\up}{\uparrow}
\newcommand{\down}{\downarrow}
  \newcommand{\dn}{\downarrow}
\renewcommand{\k}{{\bf k}}
\newcommand{\K}{{\bf K}}
\newcommand{\p}{{\bf p}}

\newcommand{\q}{{\bf q}}

\newcommand{\ek}{\epsilon_{\k}}

\newcommand{\nn}{\nonumber}
\newcommand{\beq}{\begin{equation}}
\newcommand{\eeq}{\end{equation}}

\begin{document}
\title{Medium-enhanced polaron repulsion in a dilute Bose mixture}

\author{Jesper Levinsen}
\affiliation{School of Physics and Astronomy, Monash University, Victoria 3800, Australia}
\affiliation{ARC Centre of Excellence in Future Low-Energy Electronics Technologies, Monash University, Victoria 3800, Australia}

\author{Olivier Bleu}
\affiliation{School of Physics and Astronomy, Monash University, Victoria 3800, Australia}
\affiliation{ARC Centre of Excellence in Future Low-Energy Electronics Technologies, Monash University, Victoria 3800, Australia}

\author{Meera M. Parish}
\affiliation{School of Physics and Astronomy, Monash University, Victoria 3800, Australia}
\affiliation{ARC Centre of Excellence in Future Low-Energy Electronics Technologies, Monash University, Victoria 3800, Australia}

\date{\today}

\begin{abstract}
We investigate the fundamental problem of a small density of bosonic impurities immersed in a dilute Bose gas at zero temperature. Using a rigorous perturbative expansion, we show that the presence of the surrounding medium enhances the repulsion between dressed bosonic impurities (polarons) in the regime of weak interactions. Crucially, this differs from prevailing theories based on Landau quasiparticles, which neglect the possibility of quantum degenerate impurities and predict an exchange-induced attraction. We furthermore show that the polaron-polaron interactions are strongly modified if the medium chemical potential rather than the density is held fixed, such that the medium-induced attraction between thermal impurities becomes twice the expected Landau effective interaction. Our work provides a possible explanation for the differing signs of the polaron-polaron interactions observed in experiments across cold atomic gases and two-dimensional semiconductors, and it has important implications for theories of quasiparticles and quantum mixtures in general.
\end{abstract}

\maketitle

The concept of the polaron quasiparticle---where a mobile impurity is modified by a surrounding quantum medium---is ubiquitous in physics, with relevance to systems ranging from helium mixtures~\cite{Bardeen1967} and electrons in solids~\cite{LandauPekar}, to the inner crust of neutron stars~\cite{Nemeth1968,Baym1971}. Crucially, it provides a powerful approach to describing quantum many-body systems, allowing the prohibitively many degrees of freedom to be reduced to the simpler problem of understanding the behavior of a few polaron quasiparticles~\cite{BaymPethick1991book}. Here, one simply requires key quasiparticle properties, such as the polaron energy $E_{\rm pol}$ and the polaron-polaron interaction strength $F$, to determine the physics. For instance, for bosonic impurity particles of density $n_\down$ in a medium of density $n_\up$, the energy density of the system has the form~\cite{Scazza2022}:
\begin{align}
    \mathcal{E}(n_\up,n_\down)= \mathcal{E}_0 +E_{\mathrm{pol}} n_\down+\frac12 F n_\down^2 \, ,
\label{eq:energycanonical}
\end{align}
where $\mathcal{E}_0$ is the medium energy density in the absence of impurities. Equation~\eqref{eq:energycanonical} is an exact expansion in the limit $n_\down \to 0$, and is always expected to hold for weakly interacting polaron quasiparticles~\cite{BaymPethick1991book}.

Cold-atom quantum simulators offer an exceptionally clean platform for studying polaron quasiparticles, since there is the possibility of creating highly population-imbalanced mixtures of any quantum statistics~\cite{Chevy2010,Massignan_RPP2014,Scazza2022}. This has led to the realization of both Fermi~\cite{Schirotzek2009,Nascimbene2009,Kohstall2012,Koschorreck2012,Zhang2012,Wenz2013,Cetina2015,Ong2015,Cetina2016,Scazza2017,Oppong2019,Yan2019a,Ness2020,Fritsche2021,vivanco2023} and Bose polarons~\cite{Hu2016,Jorgensen2016,Yan2019,Skou2021,etrych2024}, corresponding to impurities in degenerate Fermi and Bose gases, respectively. In parallel to these developments, advances in material science have led to precision measurements of Fermi and Bose polarons in atomically thin semiconductors, where optically excited bosons, i.e., excitons and exciton polaritons, are introduced into a gas of electrons or  polaritons~\cite{Takemura_NaturePhys2014,Sidler_NatPhys_2017,Tan_PRX20,Muir2022,Tan_PRX2023,Huang2023}. As a result of this concerted effort, there is now consensus on most single-polaron properties such as the polaron energy $E_{\mathrm{pol}}$~\cite{Massignan_RPP2014,Scazza2022}.

However, this is far from the case beyond the single-impurity limit, where, for bosonic impurities, even the sign of the medium-induced interactions in $F$ is under debate. The formalism from Landau's Fermi liquid theory has been used to argue that the sign of the effective quasiparticle interaction depends on the impurity statistics, with attractive (repulsive) interactions for bosonic (fermionic) impurities~\cite{Yu2010,YuPethick2012,Camacho2018}. Conversely, arguments based on phase-space filling (i.e., the competition between impurities for medium particles) find that the interactions between bosonic impurities are repulsive (attractive) depending on whether one is looking at the attractive (repulsive) polaron branch~\cite{Tan_PRX20,Muir2022,Tan_PRX2023}. Experimentally, there is also no consensus on the sign: For bosonic impurities, polaron interactions have been found to be attractive in cold atoms~\cite{DeSalvo2019,Baroni2024} and repulsive in two-dimensional (2D) semiconductors~\cite{Tan_PRX20,Muir2022,Tan_PRX2023}.

In this Letter, we reveal a new type of medium-induced interaction between bosonic impurities, which we show is the leading order effect for weak impurity-medium interactions. For concreteness, we consider the case of bosonic impurities in a weakly interacting Bose-Einstein condensate (BEC), corresponding to the highly spin-imbalanced limit of a two-component atomic Bose mixture at zero temperature. By performing a rigorous perturbative expansion of the ground-state energy, we show that the exchange of Bogoliubov modes enhances the existing repulsion between impurities, while exchange-induced attraction occurs at higher order in the impurity-medium interaction strength. 

We furthermore show that the polaron interactions depend strongly on whether the bosonic polarons are quantum degenerate, and---similarly to the situation in Fermi mixtures~\cite{Mora2010}---on whether the medium density or chemical potential is held fixed. Remarkably, our results are consistent with the observed sign of the polaron interactions in the stable attractive branch in all the recent experiments across the cold-atom and semiconductor platforms that feature bosonic impurities~\cite{DeSalvo2019,Tan_PRX20,Muir2022,Tan_PRX2023,Baroni2024}. We identify the key factors behind the qualitatively different results in experiment as being whether the impurities are dynamically injected with distinct~\cite{Baroni2024} or identical momenta~\cite{Tan_PRX20,Muir2022,Tan_PRX2023}, or whether the experiment is instead in thermal equilibrium at a fixed chemical potential~\cite{DeSalvo2019}.

\paragraph{Dilute Bose mixture.---} The starting point of our analysis is the energy density of a dilute mixture of two types of bosons ($\sigma=\up,\down$) at zero temperature~\cite{Larsen63}:
\begin{align}
    \mathcal{E} =&\,\frac12 \frac{4\pi a_{\up\up}n_\up^2}m+\frac12 \frac{4\pi a_{\down\down}n_\down^2}m+ \frac{4\pi a_{\up\down}n_\up n_\down}m  \nn \\ & 
    +
    \frac{256\sqrt{\pi}(a_{\up\up}n_\up)^{5/2}}{15m}f\left(\frac{a_{\up\down}^2}{a_{\up\up}a_{\down\down}},\frac{a_{\down\down} n_\down}{a_{\up\up} n_\up}\right)
    \, ,
\label{eq:Larsen}
\end{align}
where $f(x,y)=\sum_\pm \tfrac{(1+y\pm\sqrt{(1-y)^2+4xy})^{5/2}}{4\sqrt2}$, and we work in units where $\hbar=1$. In Eq.~\eqref{eq:Larsen}, the first line corresponds to the standard mean-field contribution in terms of the densities $n_\sigma$ and interparticle scattering lengths $a_{\sigma\sigma'}$ (we take $a_{\up\up},a_{\down\down}>0$ for stability, as well as $n_\sigma a_{\sigma\sigma}^3\ll1$), while the second line describes quantum fluctuations. For simplicity, we consider bosons of equal mass $m$, but the extension to arbitrary mass ratio yields qualitatively similar results (see the Supplemental Material~\cite{supmat}). Equation~\eqref{eq:Larsen}---initially derived by Larsen~\cite{Larsen63} by extending Bogoliubov theory to a two-component Bose gas---recently formed the basis of Petrov's seminal work~\cite{Petrov2015}, which predicted that a liquid-like droplet phase can be stabilized purely by quantum fluctuations in the regime $a_{\up\down}\lesssim -\sqrt{a_{\up\up}a_{\down\down}}$, where the mixture is unstable to collapse within a mean-field description. This exotic phase has since been observed in several experiments~\cite{Chomaz2016,Ferrier-Barbut2016,Cabrera2018}.

In the context of polarons and their interactions, we consider a small population of $\down$ bosons and an arbitrary sign of $a_{\up\down}$. Expanding Eq.~\eqref{eq:Larsen} to second order in $n_\down$, we find that it takes precisely the general form in Eq.~\eqref{eq:energycanonical}. Here, $\mathcal{E}_0$ coincides with the usual energy density of a weakly interacting Bose gas up to and including Lee-Huang-Yang quantum fluctuations~\cite{LHY1959,fetterbook}. Likewise, 
\begin{align}
    E_{\mathrm{pol}} &=\frac{4\pi a_{\up\down}n_\up}m\left[1+\frac{8\sqrt{2}}{3\pi}\frac{a_{\up\down}}{\xi}\right]\, ,
\label{eq:Epol}
\end{align}
exactly matches the perturbation theory results for the Bose polaron in Refs.~\cite{Novikov2009,Casteels2014,Christensen2015} up to order $a_{\up\down}^2$, with the BEC healing length $\xi=1/\sqrt{8\pi a_{\up\up}n_\up}$. 

For the polaron interaction strength, we obtain within Bogoliubov theory
\begin{align}\label{eq:gpol1}
    F&=\frac{4\pi a_{\down\down}}{m}\left[1+\frac{16\sqrt{2}}{3\pi}\frac{a_{\up\down}^2}{a_{\up\up}\xi}\left(1-\frac14
    \frac{a_{\up\down}^2}{a_{\up\up}a_{\down\down}}\right)\right]
    \,.
\end{align}
We see that $F$ contains three terms: The first is the bare interaction between impurities, while the other two are interactions mediated by the medium. Remarkably, we find that the medium-induced interaction is \textit{repulsive} at leading order in the interparticle interaction $a_{\up\down}$, and that this corresponds to an enhancement of the bare repulsion. Moreover, the total induced interaction remains repulsive whenever the mixture is stable towards collapse or phase separation, i.e., when $a_{\up\down}^2/a_{\up\up}a_{\down\down}\lesssim1$. Indeed, according to Eq.~\eqref{eq:gpol1}, the condition for the induced interaction in $F$ to be repulsive is weaker, i.e., $a_{\up\down}^2/a_{\up\up}a_{\down\down}\lesssim4$, due to a factor of 4 statistical enhancement arising from bosonic impurities being scattered into and out of the zero-momentum state. As expected, the induced interaction becomes infinite when $a_{\up\up}\to0$, a consequence of the infinite compressibility of the ideal Bose gas. 

This result differs strongly from existing calculations for non-degenerate (thermal) impurities based on Landau's effective interaction~\cite{Camacho2018}, which instead find
\begin{align} \label{eq:gpol2}
    F_\mathrm{thermal} & =\frac{4\pi a_{\down\down}}{m}\left[1-\frac{a_{\up\down}^2}{a_{\up\up}a_{\down\down}}\right]\,,
\end{align}
in the limit of weak interspecies interactions. In this case, the lowest order correction to the direct impurity interaction is independent of both $a_{\down\down}$ and $n_\up$, and it is found to be \textit{attractive} rather than repulsive. Note that this was erroneously claimed to apply also to condensed impurities~\cite{PethickComment}, which contradicts Bogoliubov theory.

\begin{figure}[tbp]    
\centering
\includegraphics[width=0.9\linewidth]{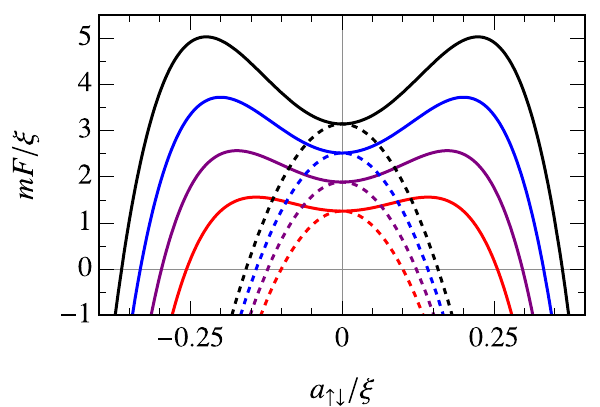}
\caption{Polaron interaction strength~\eqref{eq:gpol1} for degenerate impurities (solid) compared with the result~\eqref{eq:gpol2} for thermal impurities (dashed). From top to bottom, we have $a_{\down\down}/\xi=0.25,\,0.2,\,0.15$, and $0.1$, and for all lines $a_{\up\up}/\xi=0.1$.}
\label{fig:Fn}
\end{figure}

Figure~\ref{fig:Fn} illustrates the strong qualitative differences between the polaron interaction strengths for degenerate impurities [Eq.~\eqref{eq:gpol1}] and for thermal impurities [Eq.~\eqref{eq:gpol2}]. Not only do we see that the medium can strongly enhance the repulsion between degenerate bosonic impurities, but the critical interaction strength beyond which the interaction turns attractive is strongly modified and the attraction for larger $|a_{\up\down}/\xi|$ is significantly reduced. Since the polaron interaction is a key parameter of the quasiparticle theory, it is crucially important to understand the origin of this strong qualitative difference between degenerate and non-degenerate impurities. This is particularly pertinent in light of the discrepancies identified between different experimental platforms and protocols.

\paragraph{Two-impurity wave function approach.---}
As we now demonstrate, the medium-enhanced repulsion in Eq.~\eqref{eq:gpol1} emerges naturally when we consider the energy of two impurities in the weakly interacting regime. To perform a rigorous perturbative expansion in $a_{\up\down}$, we consider impurity-boson interaction terms beyond those included in Bogoliubov theory. Thus, we employ the Hamiltonian
\begin{align}
    \hat{H} & = 
    \sum_{\k\neq 0} E_\k \beta^\dag_\k \beta_\k + \sum_\k(\ek+ g_{\up\down} n_\up) c_\k^\dag c_\k  \nn \\ 
   & + \frac{g_{\up\down}}{V} \sum_{\p;\, \q \neq 0} c^\dag_{\p-\q} c_\p  \Big[\sqrt{N_\up} (b_{-\q} + b^\dag_{\q})  + \sum_{\k\neq\q, 0} b^\dag_\k b_{\k-\q} \Big] \nn \\
     & + \frac{g_{\down\down}}{2V} \sum_{\k\k'\q} c^\dag_\k c^\dag_{\k'} c_{\k'+\q} c_{\k-\q} \, ,
    \label{eq:HamFrohlich}
\end{align}
which is taken to be relative to the energy of the weakly interacting $\up$ Bose gas with particle number $N_\up$ in a volume $V$ (giving the density $n_\up = N_\up/V$). Here, $c^\dag_\k$ creates a spin-$\down$ impurity at momentum $\k$ and with kinetic energy $\epsilon_{\k} = |\k|^2/2m \equiv k^2/2m$, while $\beta_\k$, $\beta^\dag_\k$ are the usual $\k\neq 0$ Bogoliubov operators for the medium particles with dispersion $E_\k = \sqrt{\epsilon_{\k}(\epsilon_{\k} + 2 g_{\up\up} n_\up)}$. The bare spin-$\up$ bosons are related to the Bogoliubov modes via $b_\k = u_\k \beta_\k - v_\k \beta^\dag_{-\k}$, where $u_\k = \sqrt{(\ek + g_{\up\up} n_\up + E_\k)/2E_\k}$ and $v_\k = \sqrt{(\ek + g_{\up\up} n_\up - E_\k)/2E_\k}$. We have also introduced the interaction coefficients $g_{\sigma\sigma'}=4\pi a_{\sigma\sigma'}/m$.

To model the two-impurity problem, we introduce the variational wave function:
\begin{align}
    \ket{\Psi} & = 
    \alpha_0 c^\dag_0 c^\dag_0 \ket{\Phi} + \sum_\k \left(\alpha_\k c^\dag_\k \beta^\dag_{-\k} c^\dag_0  
    + 
    \gamma_\k c^\dag_\k c^\dag_{-\k} \right) \ket{\Phi} \nn \\ &     + \sum_{\k_1\k_2} \left(\eta_{\k_1\k_2} \, c^\dag_{\k_1+\k_2}   c^\dag_0
    + 
    \alpha_{\k_1\k_2} c^\dag_{\k_1} 
    c^\dag_{\k_2} 
    \right) \beta^\dag_{-\k_1} \beta^\dag_{-\k_2} \ket{\Phi} \nn \\
    & + \sum_{\k_1\k_2} \gamma_{\k_1\k_2} c^\dag_{\k_1} \beta^\dag_{-\k_1-\k_2} c^\dag_{\k_2} \ket{\Phi} \, ,
    \label{eq:psi}
\end{align}
which only includes the terms generated by applying the Hamiltonian twice to the ``bare'' state $c^\dag_0 c^\dag_0 \ket{\Phi}$, where $\ket{\Phi}$ corresponds to the $\up$ BEC, such that $\beta_\k\ket{\Phi} = 0$. Note that the sums exclude zero momentum and we neglect terms that vanish when the gas parameter $n_\up a_{\up\up}^3 \to 0$.

From Eq.~\eqref{eq:psi} we obtain the associated equations of motion by taking $\partial_{\lambda^*}\expval{(\hat H-E)}{\Psi}=0$ with $\lambda$ any of the amplitudes $\alpha_0,\alpha_\k,\gamma_\k,\eta_{\k_1\k_2},\alpha_{\k_1,\k_2}, \gamma_{\k_1\k_2}$. This procedure yields a set of linear equations for the expansion coefficients (for details, see \cite{supmat}). To obtain the energy $E$ in the perturbative limit, we manipulate these to have the form $E\alpha_0=\Sigma(E)\alpha_0$ with $\Sigma(E)$ the two-impurity self energy, and then perform an expansion in powers of $g_{\up\down}$, keeping only the lowest order terms in $g_{\sigma\sigma}$. This procedure yields an energy of the form $E = 2E_{\mathrm{pol}} + F/V$, with a polaron energy $E_{\mathrm{pol}}$ that matches the second-order result in Eq.~\eqref{eq:Epol} and even the third-order perturbative result of Ref.~\cite{Christensen2015}, thus demonstrating that we can accurately describe the behavior beyond Bogoliubov theory (i.e., beyond the Fr{\"o}hlich model~\cite{GrusdtDemler2015} in the polaron limit). In addition, at leading order in the gas parameter, we find that the polaron interaction constant is~\cite{supmat}
\begin{align}
    F =& g_{\down\down}\left[1 + 2 g_{\up\down}^2 n_\up \frac{1}{V}\sum_{\k} \frac{W^2_\k}{\ek + E_\k} \left(\frac{1}{\ek + E_\k} + \frac{1}{\ek} \right)\right] \nn \\ &- 2 g_{\up\down}^4 n_\up^2 \frac{1}{V}\sum_\k \frac{W_\k^4}{(\ek + E_\k)^2} \left(\frac{1}{\ek}+\frac{1}{E_\k}+\frac1{\ek+E_\k}\right) 
   \, ,
    \label{eq:interWF}
\end{align}
where $W_\k \equiv u_\k - v_\k = \sqrt{\epsilon_{\k}/E_\k}$. Taking the continuum limit and evaluating the integrals, we exactly recover the polaron interaction strength in Eq.~\eqref{eq:gpol1}. This agreement shows that Bogoliubov theory captures the leading order terms in $F$ up to order $a_{\up\down}^4$. Moreover, it confirms that the behavior of a highly spin-imbalanced Bose mixture is determined by the properties of two polaron quasiparticles, and does not require a macroscopic number of impurities.

\paragraph{Contributions beyond Bogoliubov theory.---} The $\gamma_{\k_1\k_2}$ term of our two-impurity wave function in Eq.~\eqref{eq:psi} goes beyond Bogoliubov theory since it can describe processes where a single Bogoliubov excitation is repeatedly exchanged between the two impurities. In particular, an infinite number of such exchanges gives rise to Efimov trimers~\cite{Naidon2018}, which involve an additional three-body length scale~\cite{Efimov1971,Braaten2006,Naidon2017}. This is in addition to the Efimov physics that has already been identified in the single-polaron case~\cite{Levinsen2015PRL,Yoshida2018,Christianen2022}. We find that the repeated-exchange process yields the leading order correction to $F$~\cite{supmat}:
\begin{equation}
    \Delta F= -\frac{4\pi a_{\up\down}}{m} \left( \frac{16\pi}{3}-8\sqrt{3} \right)n_\up a_{\up\down}^3 \ln(\xi/a^*)  \, ,
    \label{eq:DeltaF}
\end{equation}
where $a^* = \max(|a_{\up\down}|,a_{\sigma\sigma})$, which clearly resembles the beyond-Lee-Huang-Yang correction to the energy density of a weakly interacting single-component Bose gas~\cite{Wu1959}. However, crucially, this is higher order in the gas parameter than the attractive $n_\up^2 a_{\up\down}^4\xi^3$ contribution that has already been identified, and thus three-body (Efimov) physics is suppressed in this limit.

Another contribution beyond Bogoliubov theory is the phase-space filling effect---identified in the context of 2D semiconductors in Refs.~\cite{Tan_PRX20,Muir2022,Tan_PRX2023}. This effect would lead to odd powers of $a_{\up\down}$ in $F$, since it is repulsive when $a_{\up\down}<0$ and attractive when $a_{\up\down}>0$. However, at fixed density $n_\up$ there are no terms of order $n_\up a_{\up\down}^3\xi$ (see, also,~\cite{supmat}) and thus phase-space filling is a higher-order effect of at least $O(a_{\up\down}^5)$, beyond what we consider. Effects at stronger interactions can naturally be investigated using quantum Monte Carlo methods~\cite{Camacho_Guardian2018-prl,Ardila2022}.

\begin{figure}[tbp]    
\centering
\includegraphics[width=.95\linewidth]{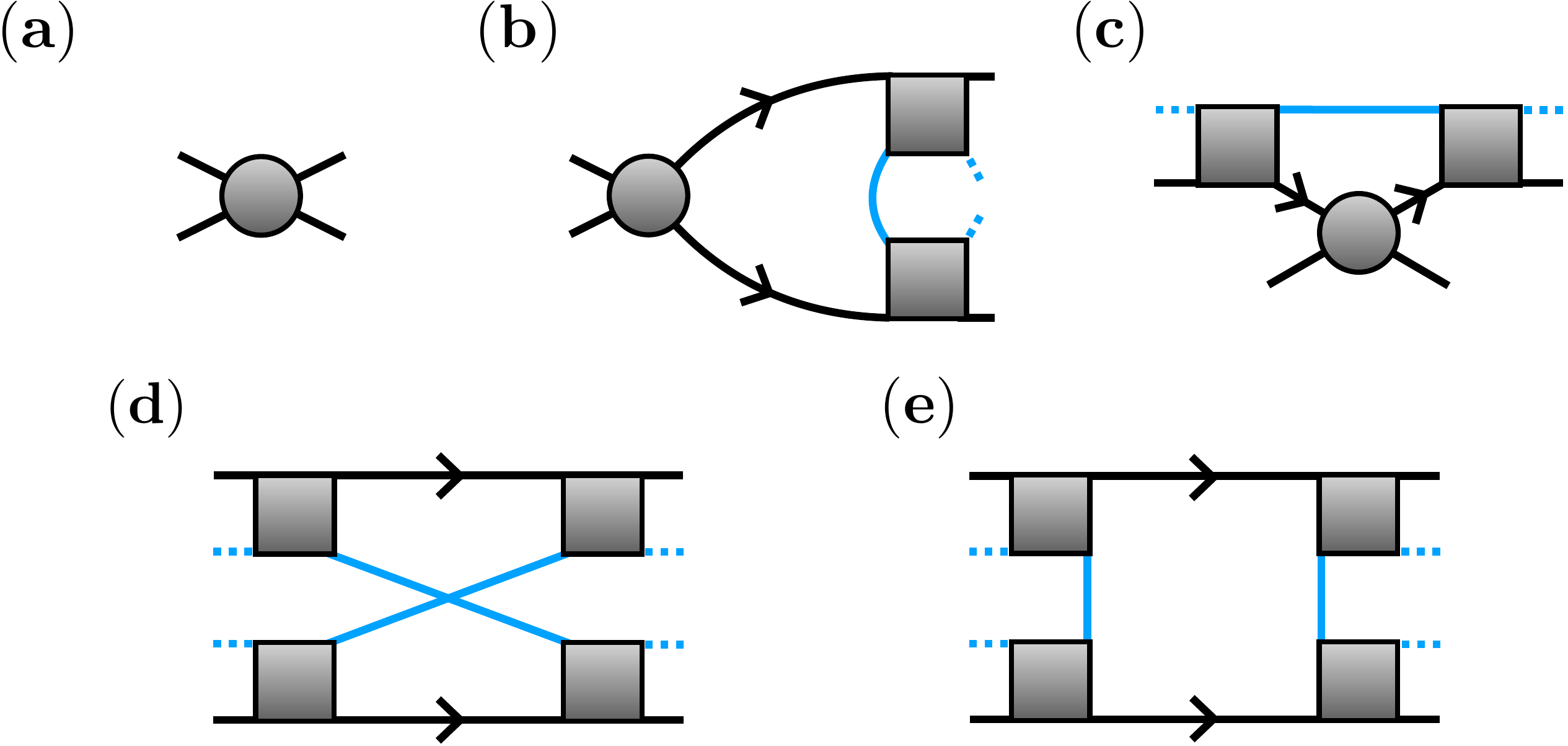}
\caption{Leading-order diagrams for the polaron interactions. The first line depicts the direct interaction in (a) and its medium-enhanced corrections (b,c), whereas the diagrams in the second line are independent of the direct interactions. The black lines denote impurity propagators, while the blue solid and dotted lines are the Bogoliubov excitations and condensate lines of the majority particles, respectively. Squares correspond to the impurity-medium interaction $g_{\up\down}$, and the circles to the direct impurity interaction $g_{\down\down}$.}
\label{fig:diagrams}
\end{figure}

\begin{figure}[tbp]    
\centering
\includegraphics[width=.82\linewidth]{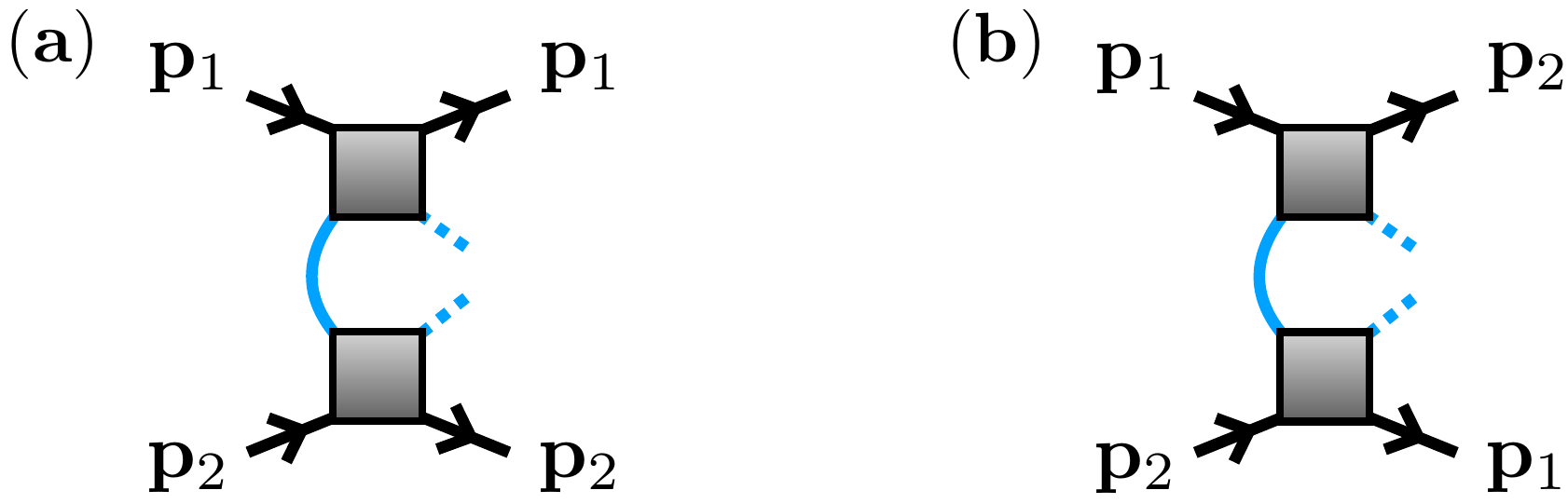}
\caption{(a) Hartree and (b) exchange interaction at order $a_{\up\down}^2$. When $\p_1 = \p_2$, as is the case for degenerate impurities, these processes coincide and do not contribute to the interaction $F$ in Eq.~\eqref{eq:energycanonical}.
}
\label{fig:diagrams2}
\end{figure}

\paragraph{Diagrammatic interpretation of polaron repulsion.---} To gain further insight into our results, we now consider a diagrammatic formulation of the polaron interaction strength $F$. Here, our perturbative result in Eq.~\eqref{eq:interWF} corresponds to the diagrams shown in Fig.~\ref{fig:diagrams}~\cite{supmat}. At first glance, panels (b-e) appear to be higher order than the $O(g_{\up\down}^2)$ exchange process in Fig.~\ref{fig:diagrams2}(b) that leads to the Landau effective interaction for thermal impurities, $F_\mathrm{th}$ in Eq.~\eqref{eq:gpol2}~\cite{Camacho2018}. However, this process does not contribute to the medium-induced interactions between polarons at equal momenta, since in this case there is no concept of exchange. Indeed, it is already known~\cite{Yu2010} that the Hartree interaction in Fig.~\ref{fig:diagrams2}(a) does not contribute to polaron interactions at fixed medium density. While the exchange process is distinct from the Hartree one when the two impurity momenta are different (i.e., $\p_1 \neq \p_2$ in Fig.~\ref{fig:diagrams2}), this is not the case for degenerate polarons where $\p_1 = \p_2$. Therefore, the diagrams in Fig.~\ref{fig:diagrams2} (which are now disconnected) do \textit{not} contribute to the medium-induced interactions between condensed polarons, e.g., in the ground state. 

Instead, the leading order diagrams must involve the scattering of impurities into finite-momentum states, as seen in Fig.~\ref{fig:diagrams}. The top row of diagrams corresponds to the first line of Eq.~\eqref{eq:interWF} and yields the medium-enhanced repulsion. Figure~\ref{fig:diagrams}(b) features the exchange of a single Bogoliubov mode, similar to Fig.~\ref{fig:diagrams2}, but in this case it involves the scattering of impurities by the direct interaction $g_{\down\down}$. The diagram in Fig.~\ref{fig:diagrams}(c), in particular, can be interpreted as Bose-enhanced repulsion arising from the scattered part (the dressing cloud) of the polaron. Here, the direct interaction $g_{\down\down}$ is sandwiched between processes that scatter an impurity into and out of the zero-momentum state. The second impurity therefore interacts with the dressing cloud of the first  and, consequently, this diagram contributes precisely $2g_{\down\down}(1-Z)$, where the factor of 2 arises from Bose statistics and $Z$ is the single-polaron residue~\cite{supmat}. 

The diagrams that only involve the impurity-medium interactions are shown in Fig.~\ref{fig:diagrams}(d,e), with their explicit expressions given in the second line of Eq.~\eqref{eq:interWF}. Thus, we see that the lowest order medium-induced attraction requires the exchange of two Bogoliubov modes rather than one, such that the impurities propagate at finite momentum in the intermediate state. On the other hand, the diagrams that lead to $\Delta F$ in Eq.~\eqref{eq:DeltaF} consist of standard three-body processes~\cite{Naidon2017} involving the exchange of a single Bogoliubov mode---see~\cite{supmat} for details.

\paragraph{Role of constraints on the medium response.---} Thus far, we have worked with a fixed medium density, as encapsulated in the energy density $\mathcal{E}$ in Eq.~\eqref{eq:energycanonical}. Here the polaron interactions are simply given by $F = \left.\pdv{^2\mathcal{E}}{n_\down^2}\right|_{n_\down \to 0}$. However, it was previously shown for highly imbalanced Fermi-Fermi mixtures that different constraints on the medium significantly affect the apparent polaron interactions, with vanishing interactions at fixed medium chemical potential~\cite{Mora2010}, and a repulsive $F$ for fixed medium density due to Pauli exchange~\cite{Mora2010,Yu2010}. 

Similarly, we find that if we fix the chemical potential $\mu_\up$ of the $\up$ bosons and consider the free energy density $\Omega(\mu_\up,n_\down) = \mathcal{E}(n_\up,n_\down) -\mu_\up n_\up$, then we instead obtain the effective polaron interaction
\begin{align} \label{eq:gpolOmega}
    F_\mu=\left.\pdv{^2\Omega}{n_\down^2}\right|_{n_\down =0} 
    = F - \left(\pdv{E_{\rm pol}}{n_\up}\right)^2
    \left.\pdv{n_\up}{\mu_\up}\right|_{n_\down =0}\, ,
\end{align}
where we have used the relation  $\mu_\up=\partial \mathcal{E}/\partial n_\up$ and we have required $\delta \mu_\up=0$ when varying the impurity density. Most notably, this is a general non-perturbative relation that is valid for \textit{any} medium or impurity statistics. For instance, Eq.~\eqref{eq:gpolOmega} is equivalent to the interactions obtained between bosonic impurities immersed in a Fermi gas at fixed chemical potential~\cite{Parish2021}. The additional term in Eq.~\eqref{eq:gpolOmega} simply arises from the change in the majority density in the presence of impurities, and is a Hartree interaction, as in Fig.~\ref{fig:diagrams2}(a).

For a dilute Bose mixture at zero temperature, the Hartree interaction exactly equals the exchange interaction from Landau's quasiparticle theory~\cite{Camacho2018}. Thus, for degenerate impurities, $F_\mu$ matches Eq.~\eqref{eq:gpol2} in the weak-coupling limit, i.e., the interaction between degenerate impurities at fixed $\mu_\up$ equals that between thermal impurities at fixed $n_\up$. By contrast, thermal bosonic impurities at fixed medium chemical potential will have \textit{twice} the induced attraction, i.e., at weak coupling:
\begin{align} \label{eq:gpol3}
    F_{\mu,\mathrm{thermal}}& =\frac{4\pi a_{\down\down}}{m}\left[1-2\frac{a_{\up\down}^2}{a_{\up\up}a_{\down\down}}\right]\, .
\end{align}
This new prediction can be tested in experiments similar to Ref.~\cite{DeSalvo2019}.

For the case of fermionic impurities in the limit of vanishing momenta, the Hartree interaction exactly cancels the exchange interaction, thus implying that there are no interactions when the medium chemical potential is fixed. This is consistent with what has been obtained for Fermi-Fermi mixtures from general thermodynamic arguments~\cite{Mora2010}. On the other hand, only the exchange term in Fig.~\ref{fig:diagrams2}(b) contributes at fixed medium density,  giving a repulsive $F$ as predicted by Landau Fermi liquid theory~\cite{Yu2010,YuPethick2012}. 

\paragraph{Discussion.---} 
To conclude, we have identified a novel mechanism of medium-induced interactions between bosonic impurities, which is dominant for weak interactions when the impurities are condensed. This effect could be detected using cold atoms by transferring a small number of bosons from a BEC into another hyperfine state, similarly to Ref.~\cite{Jorgensen2016}, thus creating degenerate impurities. Our arguments only rely on the Bose statistics of the $\down$ impurities rather than the $\up$ majority particles, and thus the medium-enhanced repulsion should apply to any quantum medium. For instance, for bosonic impurities in a Fermi gas, such as in the recent experiment~\cite{Baroni2024}, the leading order polaron interactions still correspond to the diagrams in Fig.~\ref{fig:diagrams} but with Bogoliubov and condensate lines replaced by particle and hole excitations of the medium. Moreover, we can show~\cite{BleuInPrep} that such diagrams  reproduce the results from second order perturbation theory for a Bose-Fermi mixture~\cite{Viverit2002}.

Our work also provides important insights into the polaron interactions currently observed in different platforms. In particular, the recent cold-atom experiment on polaron interactions~\cite{Baroni2024} was conducted with thermal rather than condensed impurities. We thus expect attractive exchange processes to dominate, in accordance with the experimental observations. Conversely, in the semiconductor case~\cite{Tan_PRX20,Muir2022,Tan_PRX2023}, the  excitonic impurities are optically excited at zero momentum, in which case there is no attractive Landau interaction, consistent with the observed repulsion between attractive polarons. Finally, the experiment~\cite{DeSalvo2019} on condensed bosonic impurities in a Fermi gas involved a local fixed chemical potential, since the gas was in thermal equilibrium in a harmonic trap. In this case we expect an induced attraction arising from the second term in Eq.~\eqref{eq:gpolOmega} to dominate, again in agreement with the experiment. Ultimately, our results highlight the need to carefully analyze the effective constraints on the medium response in a given experimental protocol, and to distinguish between quantum degenerate and thermal impurities.

\begin{acknowledgments}
We gratefully acknowledge fruitful discussions with Cosetta Baroni, Georg Bruun, Fr{\'e}d{\'e}ric Chevy, Victor Gurarie, Francesca Maria Marchetti, Pietro Massignan, Chris Pethick, and Matteo Zaccanti.  
We acknowledge support from the Australian Research Council Centre of Excellence in Future Low-Energy Electronics Technologies (CE170100039).
JL and MMP are also supported through Australian Research Council Discovery Project DP240100569 and Future Fellowship FT200100619, respectively.
\end{acknowledgments}

\bibliography{bosemixture}

\newcommand{\Ekup}{E_{\k}}
\newcommand{\ekdown}{\epsilon_{\k\downarrow}}

\renewcommand{\theequation}{S\arabic{equation}}
\renewcommand{\thefigure}{S\arabic{figure}}
\renewcommand{\thesection}{S\arabic{section}}

\onecolumngrid

\newpage

\setcounter{equation}{0}
\setcounter{figure}{0}
\setcounter{page}{1}

\clearpage
\section{SUPPLEMENTAL MATERIAL:\\ ``MEDIUM-ENHANCED POLARON REPULSION IN A DILUTE BOSE MIXTURE''}

\begin{center}
J. Levinsen,
O. Bleu,
M. M. Parish\\
\emph{\small School of Physics and Astronomy, Monash University, Victoria 3800, Australia and}\\
\emph{\small ARC Centre of Excellence in Future Low-Energy Electronics Technologies, Monash University, Victoria 3800, Australia}
\end{center}

In this Supplemental Material, we provide additional details on the derivations of the key results of our paper, namely the analytic expression for the polaron-polaron interaction in Eq.~(4) of the main text, its integral representation in Eq.~(8), and the leading order correction to the interaction in Eq.~(9). For added generality, we also extend the results of the main text to arbitrary masses $m_\up$ and $m_\down$ of the medium and impurity particles, respectively, with the results of the main text following from taking $m_\up=m_\down=m$.

\section{Dilute Bose mixture AT ARBITRARY MASS RATIO}
For unequal masses, the energy density of the dilute Bose mixture takes the form \cite{Petrov2015}
\begin{align}
    \mathcal{E} =\frac12 g_{\up\up}n_\up^2 +\frac12  g_{\down\down}n_\down^2+  g_{\up\down}n_\up n_\down  
    +
    \frac{8 }{15 \pi^2}m_\up^{3/2}( g_{\up\up}n_\up)^{5/2}f\left(\frac{g_{\up\down}^2}{g_{\up\up}g_{\down\down}},\frac{g_{\down\down} n_\down}{g_{\up\up} n_\up},\frac{m_\dn}{m_\up}\right)\,,
 \label{eq:Larsenpetrovmass}
\end{align}
within Bogoliubov theory. The interaction coefficients are $g_{\sigma\sigma}=4\pi a_{\sigma\sigma}/{m_\sigma}$ and  $g_{\up\down}=2\pi a_{\up\down}/{m_r}$, with reduced mass $m_r=m_\up m_\down/(m_\up+m_\down)$. The dimensionless function $f\left(x,y,z\right)$ is now given by
\begin{align}
    f\left(x,y,z\right)=  \int dq\, q^2 \left[ \tilde{E}_{+}+ \tilde{E}_{-} + \frac{1+z\left(1+4xy+y^2\right)+y^2z^2}{2q^2(z^2+1)}-\frac{q^2}{4}\left(1+\frac{1}{z}\right)-\frac{1+y}{2}\right],
\end{align}
where we have introduced
\begin{align}
    \tilde{E}_\pm =\frac{q}{4\sqrt{2}z} \left[  q^2(z^2+1)+4z(y+z) \pm \sqrt{q^2(z^2-1)+4z(y-z)+64xyz^3} \right]^{1/2}.
\end{align}

As in the equal mass case discussed in the main text, the energy density of the mixture can be be expanded to second order in $n_\down$ as
\begin{align}
      \mathcal{E}(n_\up,n_\down)=    \mathcal{E}_0+E_{\rm pol} n_\down + \frac12 F n_\down^2\, .
\end{align}
Specifically, expanding $f\left(x,y,z\right)$ to second order in $y$ gives
\begin{align}
    f\left(x,y,z\right)\simeq 1+y  \frac{15 x z}{4 (z^2-1)} \left[ \frac{ z^2 \arctan (\sqrt{z^2-1})}{\sqrt{z^2-1}} -1\right]+ y^2 \left(1-\frac{x}{4}\right) \frac{15 x z^2}{8 (z^2-1)}  \left[1 + \frac{ (z^2-2) \arctan(\sqrt{z^2-1})}{\sqrt{z^2-1}}\right].
\end{align}
One can then find the polaron energy and interaction strength by identifying the coefficients in front of the different powers of $n_\dn$ in the energy density $\mathcal{E}$.

In terms of the condensate healing length and the interspecies scattering length, the polaron energy reads
\begin{align} \label{eq:Epol-m}
E_{\rm pol} =\frac{ 2\pi a_{\up\dn} n_\up}{m_r}\left[1+ \frac{2 \sqrt{2}}{\pi  (z-1)}   \left( \frac{ z^2 \arctan (\sqrt{z^2-1})}{\sqrt{z^2-1}} -1\right)\frac{a_{\up\dn}}{\xi}\right].
\end{align}
We find that the prefactor in front of the second term matches the pertubative result of Ref.~\cite{Christensen2015}, and we recover Eq.~(3) of the main text in the limit $z\rightarrow 1$. 

For the interaction strength, we obtain
\begin{align}\label{eq:interdiffmass}
F=g_{\down\down} \left[1+2\frac{g_{\up\down}^2m_{\up}^{3/2}}{g_{\up\up}}\sqrt{g_{\up\up}n_\up}  \left(1-\frac{g_{\up\down}^2}{4 g_{\up\up}g_{\down\down}}\right) A(z) \right],
\end{align}
with
\begin{align}\label{eq:Az}
A(z) =\frac{z^2}{\pi^2(z^2-1)}\left[ 1+(z^2-2) \frac{\arctan(\sqrt{z^2-1})}{\sqrt{z^2-1}}\right].
\end{align}
When $m_\up=m_\down=m$, $A(1)=\frac{4}{3\pi^2}$ and Eq.~\eqref{eq:interdiffmass} reduces to Eq.~(4) of the main text.

Several works have investigated the effective interaction potential between two heavy impurities in the limit where they are fixed~\cite{Naidon2018,Fujii2022,Drescher2023,Yegovtsev2023,Bakkali-HassaniDalibardVarenna}. In this case, no repulsion was found. This is consistent with our results, since in this case $g_{\down\down}\to0$ and only the last term in Eq.~\eqref{eq:interdiffmass} remains. In other words, the medium-induced repulsion does not apply to the case of infinitely heavy impurities. Furthermore, we note that infinitely heavy impurities are not Bose condensed, by construction, i.e., they have well defined positions and thus do not correspond to quantum degenerate impurities at zero momentum. 

\section{IMPURITY WAVE FUNCTION APPROACH}

In this section, we outline the wave function approach in more detail. We also relax the restriction of equal masses used in the main text and consider a generalized version of the Hamiltonian in Eq.~(6): 
\begin{align} \notag
    \hat{H} =& 
    \sum_{\k\neq 0} E_{\k} \beta^\dag_\k \beta_\k + \sum_\k(\epsilon_{\k\down}+ \tilde{g}_{\up\down} n_\up) c_\k^\dag c_\k + \frac{\tilde{g}_{\up\down} \sqrt{N_\up}}{V} \sum_{\p; \k\neq0}W_\k \, c^\dag_{\p+\k} c_\p \left(\beta_\k + \beta^\dag_{-\k} \right)
    + \frac{g_{\down\down}}{2V} \sum_{\k\k'\q} c^\dag_\k c^\dag_{\k'} c_{\k'+\q} c_{\k-\q} \\ \label{eq:HamSM}
   & + \frac{\tilde{g}_{\up\down}}{V} \sum_{\substack{\p\\ \k\neq0,-\q \\\q\neq0}} c^\dag_{\p+\q} c_{\p} \left[(u_\k u_{\k + \q} + v_{\k} v_{\k+\q} ) \beta^\dag_{\k} \beta_{\k+\q} - u_\k v_{\k+\q} \beta^\dag_{\k} \beta^\dag_{-\q-\k} - u_{\k+\q} v_{\k} \beta_{-\k} \beta_{\k+\q} \right]
\, ,
\end{align}
with the bare dispersions of the two components $\epsilon_{\k\sigma} = k^2/2m_\sigma$, the Bogoliubov dispersion $E_{\k} = \sqrt{\epsilon_{\k\up}(\epsilon_{\k\up} + 2 g_{\up\up} n_\up)}$, the Bogoliubov amplitudes $u_\k = \sqrt{(\epsilon_{\k\up} + g_{\up\up} n_\up + E_\k)/2E_\k}$ and $v_\k = \sqrt{(\epsilon_{\k\up} + g_{\up\up} n_\up - E_\k)/2E_\k}$, and the impurity-boson vertex function $W_\k = u_\k - v_\k = \sqrt{\epsilon_{\k\up}/E_{\k}}$. 

The intraspecies interactions are $g_{\sigma\sigma}=4\pi a_{\sigma\sigma}/{m_\sigma}$, like before, while the interspecies interactions should instead be described by the bare coefficient $\tilde{g}_{\up\down}$, which has the perturbative expansion:
\begin{equation} \label{eq:gupdn}
    \tilde{g}_{\up\down} = g_{\up\down} + g^2_{\up\down} \sum_{\k}^\Lambda \frac{1}{\epsilon_{\k\down} + \epsilon_{\k\up}} + g^3_{\up\down} \left(\sum_{\k}^\Lambda \frac{1}{\epsilon_{\k\down} + \epsilon_{\k\up}}\right)^2 + \cdots \, ,
\end{equation}
where $\Lambda$ is a UV momentum cutoff and $g_{\up\down}=2\pi a_{\up\down}/{m_r}$, with $m_r=m_\up m_\down/(m_\up+m_\down)$ the reduced mass. Note that it is sufficient to take the lowest order term  $\tilde{g}_{\up\down} \simeq g_{\up\down}$ when calculating the polaron-polaron interaction up to order $g_{\up\down}^4 \xi^3$, but in general it is necessary to consider the expansion in Eq.~\eqref{eq:gupdn}, e.g., when calculating the polaron energy or higher order terms in the polaron-polaron interaction.

For terms of order $g_{\up\down}^2$, we only need to consider the first line in Eq.~\eqref{eq:HamSM}, corresponding to the Fr\"{o}hlich model, which is contained in the Bogoliubov theory for binary Bose mixtures. However, to completely capture terms of higher order, we also need to consider the interaction in the second line that allows impurities to scatter repeatedly with the Bogoliubov modes.

\subsection{Single polaron}

It is instructive to first consider how the case of a single impurity looks. In the weakly interacting limit, we can capture the leading order terms by considering a Chevy-type wave function with up to two excitations~\cite{Levinsen2015PRL}:
\begin{equation}
    \ket{\Psi} = \alpha_0 c^\dag_0 \ket{\Phi} + \sum_\k \alpha_\k c^\dag_\k \beta^\dag_{-\k} \ket{\Phi} + \sum_{\k_1\k_2} \eta_{\k_1\k_2} \, c^\dag_{\k_1+\k_2} \beta^\dag_{-\k_1} \beta^\dag_{-\k_2} \ket{\Phi},
\end{equation}
where $\ket{\Phi}$ is the unperturbed BEC in the absence of the impurity, such that $\beta_\k \ket{\Phi} = 0$.

To obtain the equations of motion, we minimize with respect to the coefficients, i.e., $\partial_{\alpha_0^*}\expval{(\hat H-E)}{\Psi}= 0$ etc., which gives
\begin{subequations}
\begin{align} \label{eq:a0}
    E \alpha_0 = & \tilde{g}_{\up\down} n_\up \alpha_0 + \frac{\tilde{g}_{\up\down}\sqrt{N_\up}}{V} \sum_\k W_\k \alpha_\k \, -2 \frac{\tilde{g}_{\up\down}}V \sum_{\k\k'} u_{\k} v_{\k'}\eta_{\k\k'} \\ \label{eq:ak}
    E\alpha_\k = & \left(\epsilon_{\k\down} + E_\k +\tilde{g}_{\up\down} n_\up 
    \right) \alpha_\k + \frac{\tilde{g}_{\up\down}\sqrt{N_\up}}{V} W_\k \alpha_0 \, + 2\frac{\tilde{g}_{\up\down}\sqrt{N_\up}}V \sum_{\k'} W_{\k'} \eta_{\k\k'} + \frac{\tilde{g}_{\up\down}}V \sum_{\k'} (u_\k u_{\k'} + v_\k v_{\k'}) \alpha_{\k'} \\ \notag
    E \eta_{\k_1\k_2} = & \, \left(\epsilon_{\k_1 + \k_2 \down} + E_{\k_1}  + E_{\k_2} +\tilde{g}_{\up\down} n_\up \right)\eta_{\k_1\k_2} + \frac{\tilde{g}_{\up\down}\sqrt{N_\up}}{2V} \left(W_{\k_1} \alpha_{\k_2} + W_{\k_2} \alpha_{\k_1} \right) \\ \label{eq:etakk}
    & -\frac{\tilde{g}_{\up\down}}{2V} (u_{\k_1} v_{\k_2} + u_{\k_2} v_{\k_1}) \alpha_0
    + \frac{\tilde{g}_{\up\down}}V \sum_{\k'} \eta_{\k_1\k'}(u_{\k_1} u_{\k'} + v_{\k_1} v_{\k'}) + \frac{\tilde{g}_{\up\down}}V \sum_{\k'} \eta_{\k'\k_2}(u_{\k_2} u_{\k'} + v_{\k_2} v_{\k'})
    \, .
\end{align}
\end{subequations}
Rearranging Eqs.~\eqref{eq:ak} and \eqref{eq:etakk} for $\alpha_\k$ and $\eta_{\k_1\k_2}$, respectively, expanding the expressions up to second order in $\tilde{g}_{\up\down}$, and then inserting them into Eq.~\eqref{eq:a0} and dividing out $\alpha_0$,
we obtain the single-impurity energy
\begin{align} \notag
 E = &  \tilde{g}_{\up\down} n_\up + \tilde{g}_{\up\down}^2 n_\up \frac{1}{V} \sum_\k^\Lambda \frac{W_\k^2}{E-\epsilon_{\k\down} - E_\k - \tilde{g}_{\up\down} n_\up} + \tilde{g}_{\up\down}^3 n_\up \frac{1}{V^2} \sum_{\k\k'} \frac{W_{\k} W_{\k'}(u_\k u_{\k'}+v_{\k} v_{\k'})}{(E-\epsilon_{\k\down} - E_\k - \tilde{g}_{\up\down} n_\up) (E-\epsilon_{\k'\down} - E_{\k'} - \tilde{g}_{\up\down} n_\up)} \\
 & - 2 \tilde{g}_{\up\down}^3 n_\up \frac{1}{V^2} \sum_{\k\k'}\frac{W_{\k} W_{\k'}(u_\k v_{\k'}+v_{\k} u_{\k'})}{(E-\epsilon_{\k\down} - E_\k - \tilde{g}_{\up\down} n_\up) (E-\epsilon_{\k+\k'\down} - E_{\k}- E_{\k'} - \tilde{g}_{\up\down} n_\up)}  \, ,
\end{align}
where we have neglected terms that are higher order in the gas parameter $n_\up a_{\up\up}^3$. To avoid the UV divergence as the momentum cutoff $\Lambda \to \infty$, we use Eq.~\eqref{eq:gupdn} to express the bare coupling $\tilde{g}_{\up\down}$ in terms of the physical interaction $g_{\up\down}$. Expanding up to $g_{\up\down}^3$ finally gives the perturbative expression for the polaron energy
\begin{align} \notag
    E_{\rm pol} \simeq & g_{\up\down} n_\up + g_{\up\down}^2 n_\up \frac{1}{V}  \sum_\k \left[ \frac{1}{\epsilon_{\k\down} + \epsilon_{\k\up}} - \frac{W_\k^2}{\epsilon_{\k\down} + E_\k} \right] + g_{\up\down}^3 n_\up \left( \frac{1}{V}  \sum_\k \left[ \frac{1}{\epsilon_{\k\down} + \epsilon_{\k\up}} - \frac{W_\k^2}{\epsilon_{\k\down} + E_\k} \right]\right)^2 \\ \label{eq:Epol3}
    & + 2 g_{\up\down}^3 n_\up \frac{1}{V^2} \sum_{\k\k'}\frac{W_{\k} W_{\k'}}{\epsilon_{\k\down} + E_\k} 
    \left[\frac{u_{\k'} v_\k}{\epsilon_{\k'\down} + E_{\k'}} - \frac{u_\k v_{\k'}+ u_{\k'}v_{\k} }{\epsilon_{\k+\k'\down} + E_{\k}+ E_{\k'}} \right]
    \, .
\end{align}
which we find exactly matches the diagrammatic expansion in the literature~\cite{Christensen2015}, including the logarithmic contribution that appears at order $g_{\up\down}^3$. Keeping terms up to order $g_{\up\down}^2$ yields the polaron energy $E_{\rm pol}$ in Eq.~\eqref{eq:Epol-m}.

\subsection{Two polarons}
For the case of two impurities, we use the wave function from Eq.~(7) of the main text, which we reproduce here 
%
\begin{align} \nn
    \ket{\Psi} = \  & 
    \alpha_0 c^\dag_0 c^\dag_0 \ket{\Phi} + \sum_\k \alpha_\k c^\dag_\k \beta^\dag_{-\k} c^\dag_0\ket{\Phi} + 
    \sum_\k \gamma_\k c^\dag_\k c^\dag_{-\k} \ket{\Phi} 
    + \sum_{\k_1\k_2} \gamma_{\k_1\k_2} c^\dag_{\k_1} \beta^\dag_{-\k_1-\k_2} c^\dag_{\k_2} \ket{\Phi}
    \\
    & +  \sum_{\k_1\k_2} \eta_{\k_1\k_2} \, c^\dag_{\k_1+\k_2} \beta^\dag_{-\k_1} \beta^\dag_{-\k_2}  c^\dag_0\ket{\Phi} 
    + 
    \sum_{\k_1 \k_2} \alpha_{\k_1\k_2} c^\dag_{\k_1} \beta^\dag_{-\k_1} c^\dag_{\k_2} \beta^\dag_{-\k_2} \ket{\Phi}  \, ,
\end{align}
where indistinguishability and $s$-wave interactions require $\alpha_{\k_1\k_2}=\alpha_{\k_2\k_1}$, $\eta_{\k_1\k_2} = \eta_{\k_2\k_1}$, and $\gamma_{\k_1\k_2}=\gamma_{\k_2\k_1}$, as well as $\gamma_\k = \gamma_{-\k}$ and $\alpha_\k = \alpha_{-\k}$. Similarly to the single-impurity case, we can obtain coupled equations of motion 
\begin{subequations}
    \label{eq:eomSM}
\begin{align}
    E \alpha_0 = & \left(2 \tilde{g}_{\up\down} n_\up + \frac{g_{\down\down}}{V}\right) \alpha_0 + \frac{\tilde{g}_{\up\down}\sqrt{N_\up}}{V} \sum_\k W_\k \alpha_\k + \frac{g_{\down\down}}V \sum_\k \gamma_\k + \cdots 
    \,, \\ \label{eq:gammak}
    E\gamma_\k = &  \left(2\epsilon_{\k\down} + 2 \tilde{g}_{\up\down} n_\up + \frac{g_{\down\down}}V\right)\gamma_\k 
    + \frac{g_{\down\down}}V \alpha_0 + \frac{\tilde{g}_{\up\down}\sqrt{N_\up}}V W_\k \alpha_\k + \cancel{\frac{g_{\down\down}}V \sum_{\k'\neq \k} \gamma_{\k'}} \,, \\ \notag 
    E\alpha_\k = & \left(\epsilon_{\k\down} + E_\k +2\tilde{g}_{\up\down} n_\up+ 2 \frac{g_{\down\down}}V\right) \alpha_\k + \frac{\tilde{g}_{\up\down}}V \sum_{\k'} (u_\k u_{\k'} + v_\k v_{\k'}) \alpha_{\k'} + 2\frac{\tilde{g}_{\up\down}\sqrt{N_\up}}V W_\k (\alpha_0+\gamma_\k + 
    \alpha_{\k\k} + \eta_{\k,-\k})  \\ \label{eq:ak2}
    & + 2\frac{\tilde{g}_{\up\down}\sqrt{N_\up}}V \sum_{\k'} W_{\k'} (\alpha_{\k\k'}+\eta_{\k\k'}) + 2\frac{\tilde{g}_{\up\down}}V \sum_{\k'} (u_\k u_{\k+\k'} + v_\k v_{\k+\k'}) \gamma_{\k\k'}  \,, \\ \nn
   E \gamma_{\k_1\k_2} = & \left(\epsilon_{\k_1\down} + \epsilon_{\k_2\down} + E_{\k_1+\k_2}  +2\tilde{g}_{\up\down} n_\up+ \frac{g_{\down\down}}V\right) \gamma_{\k_1\k_2} \\ 
    & + \frac{\tilde{g}_{\up\down}}{2V} \left[\alpha_{\k_1}(u_{\k_1} u_{\k_1+\k_2} + v_{\k_1} v_{\k_1+\k_2}) + \alpha_{\k_2}(u_{\k_2} u_{\k_1+\k_2} + v_{\k_2} v_{\k_1+\k_2})\right] + \cdots \,,
    \\ \label{eq:akk}
    E\alpha_{\k\k} = & \left(2\epsilon_{\k\down} + 2E_{\k} +2\tilde{g}_{\up\down} n_\up+ \frac{g_{\down\down}}V\right) \alpha_{\k\k}  + \frac{\tilde{g}_{\up\down}\sqrt{N_\up}}{V} W_{\k} \alpha_{\k} + \cdots \,, \\ \label{eq:etakk2} 
    E \eta_{\k,-\k} = & \left(2E_{\k} +2\tilde{g}_{\up\down} n_\up + \frac{g_{\down\down}}V\right)\eta_{\k,-\k} + \frac{\tilde{g}_{\up\down}\sqrt{N_\up}}{V} W_{\k} \alpha_{\k} + \cdots  \, . 
\end{align}
\end{subequations}
Here we have dropped higher-order scattering processes (e.g., we have dropped terms in Eq.~\eqref{eq:gammak} that are beyond leading order in $g_{\down\down}$), as well as terms that do not contribute to the polaron-polaron interaction up to order $g_{\up\down}^4$. In particular, for the terms in the wave function that involve two Bogoliubov excitations, only $\alpha_{\k\k}$ and $\eta_{\k,-\k}$ contribute to the polaron-polaron interaction since they involve two impurities at the same momentum, thus giving rise to a Bose enhancement. 

We can now perform a series of manipulations, similar to what we did for the single-impurity case, where we successively eliminate variables such that we finally obtain an equation of the form $E \alpha_0 =\Sigma(E) \alpha_0$. The function $\Sigma$ only depends on $E$ and the parameters of the Hamiltonian, and can be identified as the two-impurity self energy. Expanding in $g_{\up\down}$ and keeping only the leading order in $g_{\sigma\sigma}$ gives the following perturbative expression for the energy of two impurities
\begin{align} 
    E &\simeq  \underbrace{2g_{\up\down} n_\up + 2g_{\up\down}^2 n_\up \frac{1}{V}  \sum_\k \left[ \frac{1}{\epsilon_{\k\down} + \epsilon_{\k\up}} - \frac{W_\k^2}{\epsilon_{\k\down} + E_\k} \right] + \cdots}_{2E_{\rm pol}} \\ \nn
    &+ \underbrace{\frac{g_{\down\down}}V \left[1 + 2 g_{\up\down}^2 n_\up 
    \frac{1}{V} \sum_{\k} \frac{W^2_\k}{\epsilon_{\k\down} + E_\k} \left(\frac{1}{\epsilon_{\k\down} + E_\k} + \frac{1}{\epsilon_{\k\down}} \right) \right]\, 
     - 2 g_{\up\down}^4 n_\up^2 \frac{1}{V^2}\sum_\k \frac{W_\k^4}{(\epsilon_{\k\down} + E_\k)^2} \left(\frac{1}{\epsilon_{\k\down}}+\frac{1}{E_\k}+\frac1{\epsilon_{\k\down}+E_\k}\right)}_{F/V} ,
\end{align}
where $E_{\rm pol}$ corresponds to the expression in Eq.~\eqref{eq:Epol3}. When the integrals are evaluated, the interaction constant $F$ matches the analytic expression in Eq.~\eqref{eq:interdiffmass} and corresponds to Eq.~(8) in the main text when the masses are equal.

There are also terms at order $g_{\up\down}^4$ which go beyond Bogoliubov theory and are neglected in our expression for $F$. However, these scale like $n_\up g_{\up\down}^4$ or $n_\up g_{\up\down}^4 \ln(\xi)$, rather than $n_\up g_{\up\down}^4 (n_\up \xi^3)$ as in Eq.~\eqref{eq:interdiffmass}, so they are in fact higher order in the perturbative expansion for a weakly interacting Bose gas. The leading order correction to $F$ arises from the $\gamma_{\k_1\k_2}$ term in the wave function, which captures three-body correlations and thus Efimov physics between two impurities and one boson. This yields the correction:
\begin{align} \nn
\Delta F \approx
& -2 g_{\up\down}^4 n_\up \frac{1}{V^2} \sum_{\k\k'} \frac{W_{\k} W_{\k'}}{(\epsilon_{\k\down}+E_\k) (\epsilon_{\k'\down}+E_{\k'})} \frac{(u_\k u_{\k + \k'} + v_\k v_{\k+\k'})(u_{\k'} u_{\k + \k'} + v_{\k'} v_{\k+\k'})}{\epsilon_{\k\down} + \epsilon_{\k'\down}+E_{\k+\k'}}  \\
& + 2 g_{\up\down}^4 n_\up \frac{1}{V^2} \sum_\k \frac{W_{\k}^2}{(\epsilon_{\k\down}+E_\k)^2} 
\sum_{\k'} \left[ \frac{1}{\epsilon_{\k'\down} + \epsilon_{\k'\up}} - \frac{1}{\epsilon_{\k\down} + \epsilon_{\k'\down} + E_{\k+\k'}} \right] \, .
\label{eq:DeltaFSM}
\end{align}
Here we can take $u_\k \to 1$ and $v_\k \to 0$, since these contribute to higher order terms in the gas parameter $n_\up a_{\up\up}^3$. In particular, we find the integral becomes logarithmically divergent at low momenta when $a_{\up\up} \to 0$, but this is cured if we keep the Bogoliubov dispersions and thus these are sufficient for extracting the leading order behavior.

Analyzing this expression further, we see that it also depends logarithmically on the UV cutoff and thus it will diverge if we send the cutoff to infinity. However, in practice, the high-momentum behavior will be modified by higher order terms in $g_{\sigma\sigma'}$, such that we have the effective UV cutoff $1/a^*$ with $a^* = \max(|a_{\up\down}|,a_{\sigma\sigma})$. Hence, for mass ratio $z = m_\down/m_\up$, we finally obtain
\begin{equation}
    \Delta F= -\frac{2\pi a_{\up\down}}{m_r} B(z) 
    n_\up a_{\up\down}^3 \ln(\xi/a^*)  \, ,
\end{equation}
where, for equal masses, $B(1) = \frac{16\pi}{3}-8\sqrt{3}$, as in the main text.

\begin{figure}[tbp]    
\centering
\includegraphics[width=.8\linewidth]{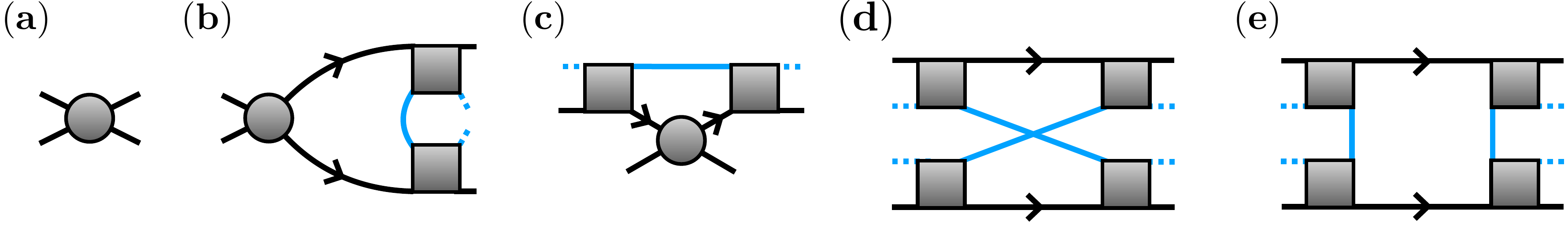}
\caption{Leading-order diagrams for the polaron-polaron interactions, reproduced from the main text.}
\label{fig:diagramsSM}
\end{figure}

\section{DIAGRAMS CONTRIBUTING TO POLARON INTERACTIONS}

In the main text, we identified the lowest-order diagrams that contribute to the polaron-polaron interactions. Here we evaluate the contribution of these diagrams to the impurity self energy using the Feynman rules for the Hamiltonian in the main text. In particular, we have the bare impurity Green's function
\begin{align}
    G_\down(\k,\omega)=\frac1{\omega-\ekdown+i0},    
\end{align}
with the small positive imaginary factor $+i0$ shifting the frequency pole slightly into the lower half of the complex plane. Likewise, at $T=0$ the majority propagators take the form~\cite{fetterbook}
\begin{subequations}
\begin{align}
    G_{11}(\k,\omega)&=\frac{u_\k^2}{\omega-E_\k+i0}-\frac{v_\k^2}{\omega+E_\k-i0}, \\
    G_{12}(\k,\omega)&=\frac{-u_\k v_\k}{\omega-E_\k+i0}-\frac{-u_\k v_\k}{\omega+E_\k-i0}, 
\end{align}
\end{subequations}
where $G_{11}$ is the normal and $G_{12}$ the anomalous propagators and we also have $G_{21}(\k,\omega)=G_{12}(\k,\omega)$ and $G_{22}(\k,\omega)=G_{11}(\k,-\omega)$.

To be concrete, we only investigate those diagrams that contribute to the interaction, and we do not consider those diagrams responsible for the polaron energy shift. These have been discussed in detail in earlier works~\cite{Novikov2009,Christensen2015}. We therefore start with the diagrams containing the direct impurity-impurity interaction. The bare impurity-impurity interaction in Fig.~\ref{fig:diagramsSM}(a) simply contributes a factor $\Sigma^{(a)}(0)=g_{\down\down}$. Here, we note that, similarly to the single-impurity problem~\cite{Christensen2015}, we can evaluate the self energy at zero energy, since energy insertions do not contribute at the order considered. Apart from this, we have two contributions. The first of these, shown in Fig.~\ref{fig:diagramsSM}(b), yields
\begin{align}
    \Sigma^{(b)}(0) 
    &=2n_\up g_{\down\down}g_{\up\down}^2 \frac iV\int \frac{d\omega}{2\pi}\sum_\k G_\down(\k,\omega)G_\down(\k,-\omega)\left[G_{11}(\k,\omega)+2G_{12}(\k,\omega)+G_{22}(\k,\omega)\right] \nn \\
    & = 4n_\up g_{\down\down}g_{\up\down}^2\frac1V\sum_\k \frac{W_\k^2 }{\Ekup+\ekdown}\frac1{2\ekdown}\,,
    \label{eq:sigmab}
\end{align}
where the factor 2 in the first line originates from the possibility of the direct interaction being on the left or right of the impurity-medium interactions, and we can replace $\tilde g_{\up\down}\to g_{\up\down}$ here and below since corrections are higher order than the diagrams considered here. Likewise, the diagram in Fig.~\ref{fig:diagramsSM}(c) contributes
\begin{align}
    \Sigma^{(c)}(0) &=2n_\up g_{\down\down}g_{\up\down}^2 \frac iV\int \frac{d\omega}{2\pi}\sum_\k G^2_\down(\k,\omega)\left[G_{11}(\k,\omega)+2G_{12}(\k,\omega)+G_{22}(\k,\omega)\right] \nn \\
    & = 2n_\up g_{\down\down}g_{\up\down}^2 \frac1V\sum_\k \frac{W_\k^2}{(\Ekup+\ekdown)^2}\,.
    \label{eq:sigmac}
\end{align}

The leading-order attraction consists of two terms. The diagram in Fig.~\ref{fig:diagramsSM}(d) yields
\begin{align}
    \Sigma^{(d)}(0) &= n_\up^2 g_{\up\down}^4 \frac iV\int \frac{d\omega}{2\pi}\sum_\k G^2_\down(\k,\omega)\left[G_{11}(\k,\omega)+2G_{12}(\k,\omega)+G_{22}(\k,\omega)\right]^2\nn \\
    & = 
    -2n_\up^2 g_{\up\down}^4 \frac1V\sum_\k \frac{W_\k^4}{(\Ekup+\ekdown)^2}\left[\frac1{\Ekup+\ekdown}+\frac1{2\Ekup}\right] \,.
    \label{eq:sigmad}
\end{align}
Finally, the diagram in Fig.~\ref{fig:diagramsSM}(e) reads
\begin{align}
    \Sigma^{(e)}(0) &= n_\up^2 g_{\up\down}^4 \frac iV\int \frac{d\omega}{2\pi}\sum_\k G_\down(\k,\omega)G_\down(\k,-\omega)\left[G_{11}(\k,\omega)+2G_{12}(\k,\omega)+G_{22}(\k,\omega)\right]^2 \nn \\
    & = 
    -2n_\up^2 g_{\up\down}^4 \frac1V\sum_\k W_\k^4\left[\frac1{(\Ekup+\ekdown)^2}\frac1{2\ekdown}+\frac1{\Ekup+\ekdown}\frac1{\Ekup-\ekdown}\left(\frac1{2\ekdown}-\frac1{2\Ekup}\right)\right] \nn \\
    & =-2n_\up^2 g_{\up\down}^4 \frac1V\sum_\k \frac{W_\k^4}{(\Ekup+\ekdown)^2}\left[\frac1{\ekdown}+\frac1{2\Ekup}\right]\,.
    \label{eq:sigmae}
\end{align}

Clearly, the sum $\Sigma^{(a)}+\Sigma^{(b)}+\Sigma^{(c)}+\Sigma^{(d)}+\Sigma^{(e)}$ precisely yields the interaction coefficient $F$ in Eq.~(8) 
of the main text. To evaluate the sum on momentum, we take the continuum limit $\frac1V\sum_\k\to\int \tfrac{d^3\k}{(2\pi)^3}$, and we find that the result matches Eq.~\eqref{eq:interdiffmass}.

It is somewhat remarkable that the self energies are simply related with $\Sigma^{(d)}+\Sigma^{(e)}=-\frac14 \frac{g_{\up\down}^2}{g_{\up\up}g_{\down\down}}(\Sigma^{(b)}+\Sigma^{(c)})$ [at the level of the integrals, this proportionality is easiest checked by relating $\partial_{1/m_\up\xi^2}(\Sigma^{(b)}+\Sigma^{(c)})$ to $\Sigma^{(d)}+\Sigma^{(e)}$, which is left as an exercise]. As for the equal mass case investigated in the main text, this proportionality is necessary to ensure the stability of finite-momentum impurities against either phase separation or collapse.

We finally remark that the polaron residue in the perturbative limit takes the form~\cite{Christensen2015}
\begin{align}
    \label{eq:residueRasmus}
    1-Z&=g_{\up\down}^2n_\up\sum_\k \frac{W_\k^2}{(\epsilon_{\k\down}+\Ekup)^2}=\frac{1}{\sqrt{2}\pi}\frac{z+1}{z-1}\left[1-\frac{\arctan(\sqrt{z^2-1})}{\sqrt{z^2-1}}\right] \frac{a_{\up\dn}^2}{a_{\up\up}\xi},
\end{align}
and thus
\begin{align}
    \Sigma^{(c)}(0)=2g_{\down\down}(1-Z),
\end{align}
at arbitrary mass ratios. This again illustrates that the process in Fig.~\ref{fig:diagramsSM}(c) corresponds to a dressing-cloud enhanced impurity-impurity scattering.

\section{THREE-BODY Diagrams}
Finally, we evaluate the contribution of three-body scattering processes to polaron-polaron interactions. Specifically, we consider processes where a medium particle is scattered out of the condensate, and repeatedly exchanged by the two impurities. Note that we do not have a contribution at $O(g_{\up\down}^3)$ since one of the impurity-medium interactions simply corresponds to a self-energy insertion on a Bogoliubov propagator, which at fixed medium density is absorbed by changing the medium chemical potential. 

\begin{figure}[tbp]    
\centering
\includegraphics[width=.52\linewidth]{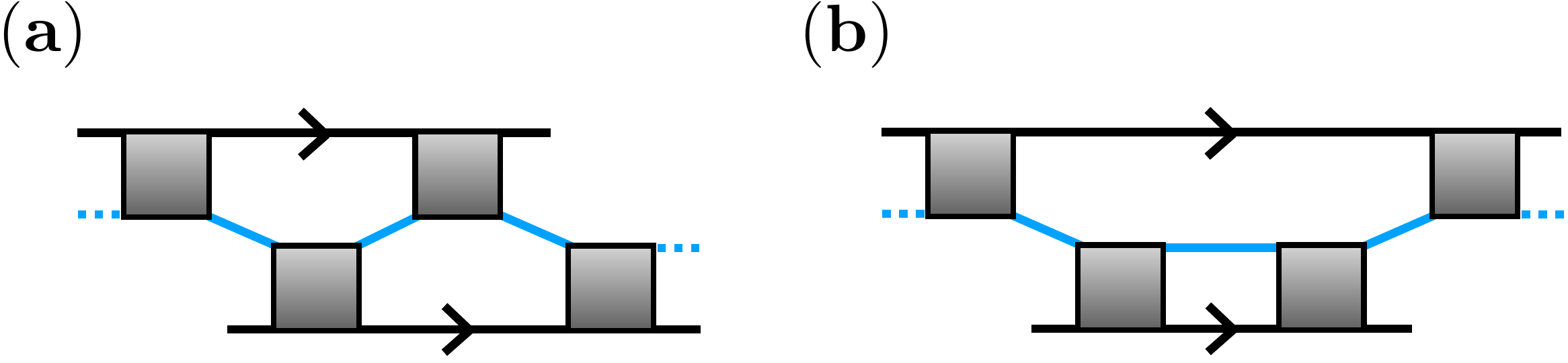}
\caption{Diagrams contributing to three-body scattering processes and Efimov physics.}
\label{fig:diagrams3body}
\end{figure}

The lowest-order non-zero contributions due to three-body processes are shown in Fig.~\ref{fig:diagrams3body}. Explicitly, the two-impurity self energy due to these two diagrams is
\begin{align}
    \Sigma^{\mathrm{(3bd)}}(0)& = 2n_0g_{\up\down}^4\frac{-1}{V^2}\int \frac{d\omega_1}{2\pi}\frac{d\omega_2}{2\pi}\sum_{\k_1\k_2}\left[G_\down(-\k_1,-\omega_1)G_\down(-\k_2,-\omega_2)G^{(0)}_\up(\k_1,\omega_1)G^{(0)}_\up(\k_2,\omega_2)G^{(0)}_\up(\k_1+\k_2,\omega_1+\omega_2)\right.
    \nn \\
    & \hspace{25mm}\left.+ G_\down(-\k_1,-\omega_1)G_\down(-\k_2,-\omega_2)\left\{G^{(0)}_\up(\k_1,\omega_1)\right\}^2\left\{G^{(0)}_\up(\k_1+\k_2,\omega_1+\omega_2)-G^{(0)}_\up(\k_2,\omega_2)\right\} \right]
    \nn \\
    &\simeq-2n_0 g_{\up\down}^4\frac1{V^2}\sum_{\k_1\k_2}\left[\frac{1}{\epsilon_{\k_1\down}+\epsilon_{\k_1\up}}\frac{1}{\epsilon_{\k_2\down}+\epsilon_{\k_2\up}}\frac1{\epsilon_{\k_1\down}+\epsilon_{\k_2\down}+\epsilon_{\k_1+\k_2\up}}\right. \nn \\ & \hspace{30mm}\left.+\frac{1}{(\epsilon_{\k_1\down}+\epsilon_{\k_1\up})^2}\left(\frac1{\epsilon_{\k_1\down}+\epsilon_{\k_2\down}+\epsilon_{\k_1+\k_2\up}}-\frac{1}{\epsilon_{\k_2\down}+\epsilon_{\k_2\up}}\right)\right].\label{eq:3bodydiag}
\end{align}
The subtracted term in the second diagram is due to the renormalization of $g_{\up\down}$. Here, we have set $E_\k\to \epsilon_{\k\up}$, $u_\k\to1$ and $v_\k\to0$ since corrections to these limits are even higher order in the gas parameter. This enables us to effectively work with the vacuum limit of the medium propagator
\begin{align}
    G^{(0)}_\up(\k,\omega)=\frac1{\omega-\epsilon_{\k\uparrow}+i0}.    
\end{align}
Comparing Eq.~\eqref{eq:3bodydiag} with Eq.~\eqref{eq:DeltaFSM} (in which we can also safely set $W_\k\to1$, $E_\k\to \epsilon_{\k\up}$, $u_\k\to1$ and $v_\k\to0$ at leading order in the gas parameter since this only enters the logarithm) we see that the variational and diagrammatic approaches again give exactly the same contribution. 

The calculation of higher-order diagrams proceed in much the same manner. Importantly, to see that such processes are subleading, we do not need to evaluate the precise form of the diagrams, we just need to estimate their order. For instance, we can estimate the order of the processes in Fig.~\ref{fig:diagrams3body} by introducing a 6-dimensional  hypermomentum $\K$ to find that the sum can be approximated as
\begin{align}
    m_{\down}^3 n_0 g_{\up\down}^4\int d^6K \frac{1}{K^6\Phi(\Omega_\K)}\sim \frac{n_0 a_{\up\down}^4 \ln(\xi/a^*)}{m_\down}\,,
\end{align}
with $\Phi(\Omega_\K)$ a function of the hyperangles. In the last step, we have evaluated the integral using the fact that it is both infrared and ultraviolet divergent. The infrared divergence is cured by properly accounting for the linear phonon dispersion in the BEC, whereas the ultraviolet cutoff is provided by retaining the energy dependence of the scattering amplitudes. This argument clearly shows that it is straightforward to estimate the order of a given diagram.

This procedure is easily generalized to higher order diagrams featuring repeated exchange of a single Bogoliubov excitation. Each time we have an additional $\up$-$\down$ interaction in a given three-body process, it adds an extra sum on a momentum and a majority propagator (the extra impurity propagator essentially vanishes when integrating over the associated frequency as in the diagrams above). Thus, at order $l\geq5$, similarly to above, we have a contribution
\begin{align}
    m_{\down}^{(l-1)} n_0 g_{\up\down}^l\int d^{3(l-2)}K \frac{1}{K^{2(l-1)}\Upsilon(\Omega_\K)}\sim \frac{n_0 a_{\up\down}^l}{m_\down} \left(\frac1{a^*}\right)^{l-4}\stackrel{\tiny{|a_{\up\down}|\gtrsim a_{\up\up}}}{\sim} \frac{n_0 a_{\up\down}^4}{m_\down}\,,
\end{align}
where $\Upsilon$ is another function of the hyperangles. Here, the order of the integral is obtained by noting that it is no longer infrared divergent when $l\geq 5$, whereas the ultraviolet divergence is again cured by treating the two-body scattering matrix appropriately. This in turn sets a typical ultraviolet momentum $K\sim (a^*)^{-1}$ which dominates the integral.

To conclude, diagrams such as those in Fig.~\ref{fig:diagrams3body} are higher order than those considered in the main text. Notably, the three-body diagrams are responsible for Efimov physics of two impurities and one majority particle. While Efimov physics is suppressed in the perturbative regime investigated in this work, it is expected that such processes become important when the $\up$-$\down$ interactions are stronger, particularly for heavy impurities in a BEC of light bosons~\cite{Naidon2018}.

\end{document}